\documentclass[pra,aps,twocolumn,epsfig,nofootinbib]{revtex4}

\usepackage{amsmath}
\usepackage{amssymb}
\usepackage[dvips]{graphicx}
\usepackage{dcolumn}
\usepackage{bm}
\usepackage[colorlinks=false,dvipdfm]{hyperref} 
\usepackage{setspace}
\usepackage{amsfonts}
\usepackage{rotating}
\usepackage{makeidx}
\usepackage{amsthm}

\newcommand{\tr}{\mbox{tr}}

\newcommand{\rvs}{ }

\begin{document}

\title{Intrinsic coherence in assisted sub-state discrimination \footnote{EPL, 117 (2017) 10013}}

\author{Fu-Lin Zhang}
\email[Corresponding author: ]{flzhang@tju.edu.cn}
\affiliation{Physics Department, School of Science, Tianjin
University, Tianjin 300072, China}

\author{Teng Wang}
\affiliation{Physics Department, School of Science, Tianjin
University, Tianjin 300072, China}

\date{\today}

\begin{abstract}
We study intrinsic coherence in the tripartite process to unambiguously discriminate two nonorthogonal states of a qubit, entangled with another one, and assisted by an auxiliary system.
The optimal success probability is found to be benefited by initial intrinsic coherence, but no extra one is required.
The transformations among different contributions of intrinsic coherence are necessary in this procedure, which increase with the overlap between the states to recognize.
Such state discrimination is a key step of the probabilistic teleportation protocol.
Entanglement of the quantum channel decreases the coherence characterizing the reliance on an ancilla.
\end{abstract}



\maketitle



%
%
%

\section{Introduction}\label{Intro}

Coherence is an essential distinction between a quantum system and a classical one.
It is the origin of nonclassical correlations in composite systems.
Many concepts have been developed to describe these correlations, such as entanglement \cite{RevModPhys.81.865}, Bell nonlocality \cite{RMP2014bell}, discord \cite{RMP2012Vedral} and Einstein-Podolsky-Rosen (EPR) steering \cite{PRL2007Steering,PRX2015SteeringRes}, etc.
They are the key factors that allow us to perform some quantum information processes with or without classical counterparts.
On the other hand, the development of quantum information science provides a perspective to reassess these correlations in the formulation of resource theory \cite{RevModPhys.81.865,PRX2015SteeringRes}.
Recently, this perspective has been used to quantify coherence \cite{PRL2014Coherence}, which gained widely attention and raised a subsequent stream of works relevant to quantum coherence \cite{PRL2014ObsCoher,PRL2015CoherEnt,PRL2016CoherDiscord,PRL2016CoherPath,PRL2016CoherMulti}.

One of the topics in these researches is the distribution of coherence in multipartite systems.
In their very recent work  \cite{PRL2016CoherMulti}, Radhakrishnan \emph{et. al.} decompose multipartite coherence into local and intrinsic parts.
The former denotes the coherence localized on individual sites, while the latter cannot be attributed to particular subsystems and is shown to equal to entanglement within the whole system.
The intrinsic coherence in a multipartite system can be further decomposed into several bipartite coherences.
These coherence contributions and their trade-off relations as quantum properties in a multipartite system have been illustrated by using some special states and many-body models in \cite{PRL2016CoherMulti}.
Here a natural question is left: What roles do they play in specific quantum information tasks?

In this work, we focus on the intrinsic coherence in the procedure to unambiguously discriminate two nonorthogonal states of a qubit entangled with another one. \rvs{
Here we show that the initial intrinsic coherence benefits the optimal success probability and no extra intrinsic coherence is required in the procedure .
In addition, the transformations among different contributions of intrinsic coherence are always necessary, which is increased by the inner product between the states to recognize.

We refer to this task as \emph{unambiguous sub-state discrimination} (USSD), to differentiate with the case with a \emph{priori} classical probabilities studied in \cite{roa2011dissonance,PhysRevA.85.022328,zhang2012assisted}.
In fact, such process exists in specific quantum information protocols, for instance the probabilistic teleportation \cite{ConcluTeleport,PRA2015RoaTele} which will be shown as a part of our results.
And the discrimination with classical probabilities corresponds to a special case of the present work.

Our basic idea to study the intrinsic coherence in  USSD comes from the researches about quantum correlations in unambiguous state discrimination with a \emph{priori} classical probabilities \cite{roa2011dissonance,PhysRevA.85.022328,zhang2012assisted}.
This process can be implemented successfully in the absence of quantum entanglement and is conjectured to be aided by quantum discord. 
But there exists a region where the optimal state discrimination can be achieved without discord \cite{zhang2012assisted}.
This prompts us to examine further the quantumness in the protocol.
The key step in such process is a joint unitary transformation between the system and an ancilla, followed by orthogonal measurements on them.
Intuitively, the quantumness is produced by the unitary operation and consumed in the measurement.
A natural candidate is the coherence in the whole system, as the orthogonal measurements, with respect to their corresponding basis, are incoherent operations \cite{PRL2015CoherEnt}.
However, the optimal state discrimination is a problem invariable under local unitary (LU) transformations.
This points to the intrinsic coherence  between the system and the ancilla, but it has already been proved to be completely unnecessary in the process \cite{zhang2012assisted}.
We argue that this confusion comes from the influence of classical probabilities on the quantumness, which is also a major reason why there are different ways to divide the quantum and classical worlds \cite{RMP2012Vedral}.

To overcome this, we replace the classical probabilities with an \emph{environment} qubit, which is prepared in an entangled pure state with the system qubit.
Then, the state discrimination becomes a tripartite process, and the intrinsic coherence drawing our attention is extended to the one among the whole tripartite system.
The details of our results are given after a brief review of tripartite intrinsic coherence.

\section{Tripartite intrinsic coherence}
For convenience in the following parts of this paper, we denote the three subsystems of a tripartite system as $S$, $C$ and $A$.
According to the results in \cite{PRL2016CoherMulti}, its total intrinsic coherence can be decomposed into bipartite coherences as
}
\begin{equation}
\mathcal{C}_{I}  \simeq  \mathcal{C}_{S:C} +  \mathcal{C}_{A:SC} \simeq  \mathcal{C}_{C:A} +  \mathcal{C}_{S:CA} \simeq  \mathcal{C}_{A:S} +  \mathcal{C}_{C:AS},
\label{CI}
\end{equation}
where $\mathcal{C}_{x:yz}$ stands for the intrinsic coherence between the bipartition of sites $x$ and $yz$ and $\mathcal{C}_{y:z}$ for the one between $y$ and $z$, with $x, y, z = \{A, S, C \}$.
The genuine tripartite coherence can be estimated by subtracting pairwise bipartite terms from the total intrinsic coherence, giving
\begin{equation}
\mathcal{C}_{I}  - \mathcal{C}_{S:C} - \mathcal{C}_{C:A} -\mathcal{C}_{A:S}  \simeq  \mathcal{C}_{C:AS}- \mathcal{C}_{C:A}-  \mathcal{C}_{S:C}.
\label{Cg}
\end{equation}
The right-hand side holds for any other permutation of the subscripts,
 and is a measure of the multipartite monogamy of coherence.


\section{Sub-state discrimination}\label{ASSD}

Consider that  Alice and Bob share an entangled two-qubit state
\begin{equation}\label{StateSC}
|\chi\rangle=\sqrt{r_+}|\xi\rangle_s |\varphi\rangle_c + \sqrt{r_-}|\bar{\xi}\rangle_s |\bar{\varphi}\rangle_c,
\end{equation}
where $\{ |\xi\rangle_s, |\bar{\xi}\rangle_s \}$ are two nonorthogonal states of the qubit of Alice (system qubit $S$),
and $\{|\varphi\rangle_c,|\bar{\varphi}\rangle_c \}$ are the ones of qubit $C$ in Bob's hand.
Suppose the overlaps $_s \langle\bar{\xi}|\xi\rangle_s=\alpha=|\alpha|e^{i \gamma_s}$, $_c \langle\bar{\varphi}|\varphi\rangle_c=\alpha_c=|\alpha_c|e^{i \gamma_c}$, thus the phase difference between the two terms of state $|\chi\rangle$ is $\gamma=\gamma_s+\gamma_c$.
The normalization condition, $\langle \chi |\chi\rangle=1$, requires that $r_{\pm}=p_{\pm}/(1+2 \sqrt{p_+ p_-}|\alpha||\alpha_c| \cos \gamma)$ with $p_{\pm} \in [0,1]$ and $p_+ + p_- =1$.
The task of Alice is to measure, with no error permitted, the two states $|\xi\rangle_s $ and $|\bar{\xi}\rangle_s$ of $S$, and collapse $C$ onto $|\varphi\rangle_s $ or $|\bar{\varphi}\rangle_s$ simultaneously.
Since $|\xi\rangle_s $ and $|\bar{\xi}\rangle_s $ are not orthogonal, the distinction may sometimes fail as the the price to pay for no error.
Therefore, Alice's measurement has three possible outcomes,  $|\xi\rangle_s $, $|\bar{\xi}\rangle_s $, and inconclusive.

The case with a \emph{priori} classical probabilities studied in \cite{roa2011dissonance,PhysRevA.85.022328,zhang2012assisted} can be realized by the reduced state of $S$ of the whole state $|\chi\rangle$ in (\ref{StateSC}) with $|\alpha_c|=0$ , and thereby it is a special case of our present results.
Similarly to \rvs{the discrimination among three states with classical probabilities \cite{PRL2012Discr}}, one can find that the phase $\gamma$ is the Berry phase associated with a closed path in parameter space ($|\chi_1\rangle \rightarrow |\chi_2\rangle \rightarrow |\chi_3\rangle\rightarrow |\chi_1\rangle $) \cite{Berry}, as $\gamma= \arg \langle\chi_1| \chi_2 \rangle +\arg \langle\chi_2| \chi_3 \rangle+\arg \langle\chi_3| \chi_1 \rangle$.
Here, $|\chi_1\rangle=|\chi(p_+=0)\rangle$, $|\chi_2\rangle=|\chi\rangle$, and $|\chi_3\rangle=|\chi(p_-=0)\rangle$.
It is invariant under the $ U(1)$ transformation $|\chi_j\rangle \Rightarrow e^{i \theta_j}|\chi_j\rangle$.
Obviously, the Berry phase is a significant factor that impacts the quantum coherence in state $|\chi\rangle$.
This impact also exists in the whole process of USSD, which is shown in the rest of this article.

To discriminate her two states unambiguously, Alice couples the system to an auxiliary qubit $A$, prepared in a known initial pure state $|k\rangle_a$.
Under a joint unitary transformation $U_{SA}$ between $S$ and $A$, the three-qubit state becomes
\begin{equation}\label{StateASC}
|\Gamma\rangle=U_{SA}|\chi\rangle |k\rangle_a =\sqrt{r_+}|\zeta_+\rangle  |\varphi\rangle_c + \sqrt{r_-}|\zeta_-\rangle |\bar{\varphi}\rangle_c,
\end{equation}
with
\begin{subequations}\label{Utrans}
\begin{align}
|\zeta_+\rangle =\bar{\alpha}_{+}\;|0\rangle_s|0\rangle_a+\alpha_+|\eta\rangle_s|1\rangle_a, \\
|\zeta_-\rangle =\bar{\alpha}_{-}\;|1\rangle_s|0\rangle_a+\alpha_-|\eta\rangle_s|1\rangle_a,
\end{align}
\end{subequations}
where $\bar{\alpha}_{\pm} = \sqrt{1-|\alpha_{\pm}|^2}$, $|\eta\rangle_s=\cos \beta |0 \rangle_s+ \sin \beta e^{i \delta} |1\rangle_s$, $\{|0\rangle_s,|1\rangle_s\}$ and $\{|0\rangle_a,|1\rangle_a\}$ are the \rvs{basis} for the system and the ancilla respectively.
The probability amplitudes $\alpha_+$ and $\alpha_-$ satisfy $\alpha_+\alpha_-^\ast=\alpha$.

Subsequently, Alice performs a von Neumann measurements on the \rvs{basis} $\{|0\rangle_a,|1\rangle_a\}$ of the ancilla.
Her discrimination is successful if the outcome is $0$, since in this case $S$ collapses to the orthogonal states $|0\rangle_s$ or $|1\rangle_s$, corresponding to $|\xi\rangle_s$ and $|\bar{\xi}\rangle_s$ respectively.
Otherwise, she fails if the outcome is $1$, as $S$ collapses to $|\eta\rangle_s$.
The success probability is given by
 \begin{eqnarray}\label{prob}
P_{\rm suc}&=& \langle \Gamma |  (\openone_s \otimes  \openone_c \otimes |0\rangle_a\langle0|  )  |\Gamma\rangle \nonumber\\
&=&r_+ (1-|\alpha_+|^2)+r_- (1-|\alpha_-|^2),
\end{eqnarray}
where $\openone_s$ and $\openone_c$ are the unit operators for qubit $S$ and $C$ respectively.
Without loss of generality, we take $p_+\leq p_-$ and denote $\tilde{\alpha}=\sqrt{p_+/p_-}$.
The optimal success probability can be analyzed as two cases:
(i) $|\alpha| < \tilde{\alpha}$, $P_{\rm suc, max}$ is attained for $|\alpha_+|=\sqrt{|\alpha|/\tilde{\alpha}}$;
(ii) $\tilde{\alpha}\leq |\alpha|\leq 1$, $P_{\rm suc, max}$ is attained for $|\alpha_+|=1$, in which Alice \rvs{ignores} the state $|\xi\rangle$ according to her \emph{priori} knowledge of the state $|\chi\rangle$.
One has
\begin{subequations}\label{twocase}
\begin{align}
&&P_{\rm suc, max} = r_+ + r_- -2 \sqrt{r_+ r_-} |\alpha|,\  {\rm for \;\; case (i)},\label{case1} \\
&&P_{\rm suc, max} = r_- (1-|\alpha|^2),\ \ \ \  \ \ \  \ \ \  \ \ \   {\rm for \;\; case (ii)}.\label{case2}
\end{align}
\end{subequations}

\begin{figure}
\includegraphics[width=7.5cm]{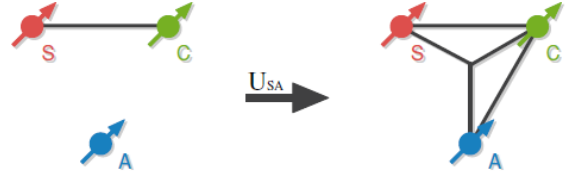} \\
\caption{(Color online) In the protocol of USSD, the joint unitary transformation $U_{SA}$ does not affect the total intrinsic coherence but only changes its distribution among the three qubits, as the coherence contribution between $S$ and $A$ remains zero.
} \label{fig1}
\end{figure}

\section{Separability between $S$ and $A$}

It is obvious that, the procedure and its optimum conditions, which are the same as the results in \cite{roa2011dissonance,zhang2012assisted}, are independent of the overlap $\alpha_c$ and the Berry phase $\gamma$.
However, the state $|\Gamma\rangle$ and its reduced state of the system and ancilla
 \begin{eqnarray}\label{rhosa}
\rho_{SA} &=& \tr_c |\Gamma\rangle \langle \Gamma | \nonumber\\
 &=& r_+ |\zeta_+\rangle \langle \zeta_+ | + r_-|\zeta_-  \rangle \langle \zeta_-| \nonumber\\
  &\ & \  + \sqrt{r_+ r_-}(\alpha_c |\zeta_+\rangle \langle \zeta_- |+ \alpha_c^\ast |\zeta_-\rangle \langle \zeta_+ |),
\end{eqnarray}
clearly depend on the two parameters.
When $|\alpha_c| =0$, the intrinsic coherence (entanglement) of $\rho_{SA}$ has been proved to be completely unnecessary for the discrimination \rvs{process} \cite{roa2011dissonance,zhang2012assisted}.
\rvs{In the following paragraph, we show that} the result holds true for USSD.
Consequently, the total intrinsic coherence keeps unchanged before and after the joint unitary transformation as shown in Fig. \ref{fig1} according to the relations in (\ref{CI}), as $U_{SA}$ does not affect the intrinsic coherence between the bipartition of sites $C$ and $AS$.
In other words, no extra intrinsic coherence beyond \rvs{the initial one in state} $|\chi\rangle$ is needed for performing USSD.

The success probability is independent of the state $|\eta\rangle_s$ in $U_{SA}$.
\rvs{
One can derive the value of $\beta$ and $\delta$ corresponding to the absence of  intrinsic coherence in $\rho_{SA}$.
To do so, we first rewrite it as
}
 \begin{eqnarray}\label{rhosaSep}
\rho_{SA}  = |\zeta_1\rangle \langle \zeta_1 |+  |\zeta_2\rangle \langle \zeta_2 | ,
\end{eqnarray}
where $|\zeta_j\rangle = q_j^+ |\zeta_+\rangle + q_j^- e^{i \gamma_j} |\zeta_-\rangle $ with the real parameters satisfying the constrains ${q_1^{\pm}}^2+{q_2^{\pm}}^2=r_{\pm}$ and $q_1^+ q_1^- e^{i \gamma_1}+q_2^+ q_2^- e^{i \gamma_2} = \sqrt{r_+ r_-}\alpha_c^\ast$.
The concurrence of the two pure states are given by
\begin{eqnarray}\label{ConZeta}
{\rm Con}(\zeta_j)&=&2|q_j^+ \alpha_+ + q_j^- \alpha_- e^{i \gamma_j}| \nonumber \\
&&  \times |q_j^+ \bar{\alpha}_+ \sin \beta e^{i \delta}-q_j^- \bar{\alpha}_- \cos \beta  e^{i \gamma_j}|. \ \ \ \ \ \ \
\end{eqnarray}
Let $\sqrt{q_{\pm}}e^{i \omega_{\pm}}=\sqrt{r_\pm} \alpha_\pm + \sqrt{r_{\mp}} \alpha_{\mp} |\alpha_c| e^{\mp i \gamma_c}$ with $q_{\pm}$ and $\omega_{\pm}$ being real.
One can find that, $q_1^{\pm}=\sqrt{r_{\pm} q_{\pm}/(1-P_{\rm suc})}$, $\gamma_1=\omega_+-\omega_-$, $q_2^{\pm}=|\alpha_{\mp}|\sqrt{r_+ r_-/(1-P_{\rm suc})}$ and $\gamma_2=\gamma_s-\pi$ is the solution to the constrains, which makes the first factor of the concurrence for $|\zeta_2\rangle$ in (\ref{ConZeta}) to be zero.
Then, the concurrence ${\rm Con}(\zeta_1)$ can be set to zero by adjusting the parameters in the second factor as
\begin{equation}\label{Con0}
\beta=\arctan \sqrt{\frac{p_- q_- (1-|\alpha_-|^2)}{p_+ q_+ (1-|\alpha_+|^2)}},\ \ \ \delta=\omega_+-\omega_-.
\end{equation}
That is, the intrinsic coherence in the two-qubit state $\rho_{SA}$ is absent when the the parameters in state $|\eta\rangle_s$ satisfy the relations in (\ref{Con0}), as it can be written as \rvs{a mixture} of two separable pure states.

\section{Intrinsic coherence}\label{IntrCoher}

From the previous analysis we learned that, the joint unitary transformation $U_{SA}$ does not affect the total intrinsic coherence but only changes its distribution among the three qubits.
The minimal amount of total intrinsic coherence required in the optimal USSD is provided by the initial state $|\chi\rangle$.
We shall investigate the relations between various contributions of intrinsic coherence and the procedure of optimal USSD in this case.

To analytically express our results, we adopt \emph{tangle} (squared concurrence) \cite{Wootters} as the measure of intrinsic coherence rather than the one based on the Jensen-Shannon divergence proposed in \cite{PRL2016CoherMulti}.
An evident advantage of our choice is that the symbols of approximately equal in (\ref{CI}) and (\ref{Cg}) can be replaced by equal signs, since the three qubits in the procedure are in a pure state \cite{Tangle}.
According to the relations in (\ref{CI}) and (\ref{Cg}), all of the nonzero intrinsic coherence in the tripartite state $|\Gamma\rangle$ can be obtained by simple addition and subtraction of the following three values,
\begin{eqnarray}\label{CIvalue}
&&\mathcal{C}_I=4 r_+ r_- \bar{\alpha}_c^2 (1-|\alpha|^2), \nonumber \\
&&\mathcal{C}_{A:SC}=4 r_+ r_- \bar{\alpha}_c^2 (|\alpha_+|^2+|\alpha_-|^2-2|\alpha|^2),  \\
&&\mathcal{C}_{g}=4 r_+ r_- \bar{\alpha}_c^2  |\bar{\alpha}_+ \alpha_- \sin \beta e^{i \delta}+\bar{\alpha}_- \alpha_+ \cos \beta|^2, \ \ \ \ \ \  \nonumber
\end{eqnarray}
where $\bar{\alpha}_c=\sqrt{1-|\alpha_c|^2}$ and $\mathcal{C}_{g}$ is the genuine tripartite coherence defined in (\ref{Cg}).
In the case of optimal discrimination, they \rvs{depend on} the values of $p_+$, $|\alpha|$, $|\alpha_c|$ and $\gamma$.

\begin{figure}
\includegraphics[width=8cm]{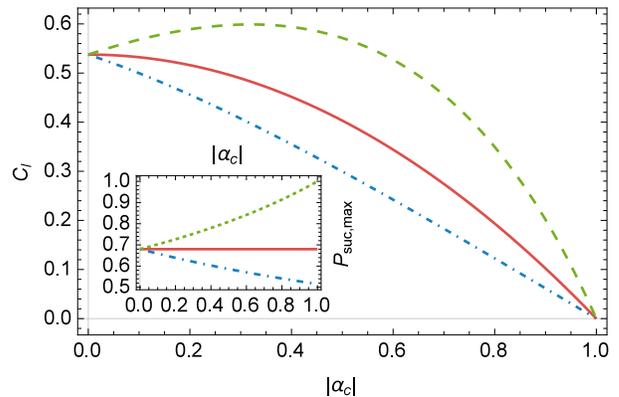} \\
\caption{(Color online) Initial intrinsic coherence between $S$ and $C$ as a function of the overlap $|\alpha_c|$ when $|\alpha|=0.4$ and $p_+=0.2$, for the Berry phase $\gamma=0$ (dot-dashed line), $\gamma=\pi/2$ (solid line), and $\gamma=\pi$ (dashed line). Inset shows the optimal success probability under the same parameters.
} \label{fig2}
\end{figure}

Let us begin with the total intrinsic coherence $\mathcal{C}_I $, \rvs{which equals the initial one between $S$ and $C$}.
In Fig. \ref{fig2}, we show the relation between $\mathcal{C}_I $ and the overlap $|\alpha_c|$, together with the behavior of the optimal success probability.
For fixed amounts of $p_+$, $|\alpha|$ and $|\alpha_c|$, $\mathcal{C}_I $ decreases with $\cos\gamma$.
Its minimum and maximum of are attained \rvs{when} the Berry phase is $\gamma=0$ and $\gamma=\pi$ respectively.
The relation between $P_{\rm suc,max}$ and $|\alpha_c|$ relies on the value of $\gamma$.
The optimal success probability decreases with $|\alpha_c|$ when $\cos \gamma >0$, but increases for $\cos \gamma <0$.
More particularly, when $\gamma =\pi$, with the \rvs{increase} of $|\alpha_c|$, it approaches to $1$.
Evidently these behaviors spring from the coherent superposition of the two terms of state $|\Gamma\rangle$ in (\ref{StateASC}).
The terms with $|1\rangle_a$ \rvs{in (\ref{Utrans})} corresponding to the probability of failure, reinforce or cancel out each other in the two cases of  $\cos \gamma >0$ and  $\cos \gamma <0$ respectively.
We can draw the conclusion that, the initial intrinsic coherence $\mathcal{C}_I $ enhances the optimal success probability for fixed $|\alpha_c|$, and determines its evolutionary trend with $|\alpha_c|$.

\begin{figure}
\centering
\includegraphics[width=8cm]{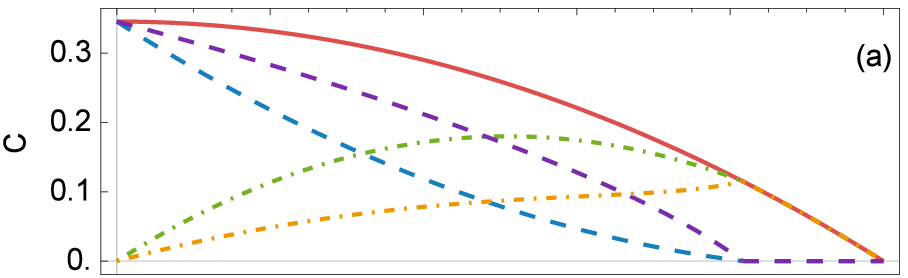}\
\includegraphics[width=8cm]{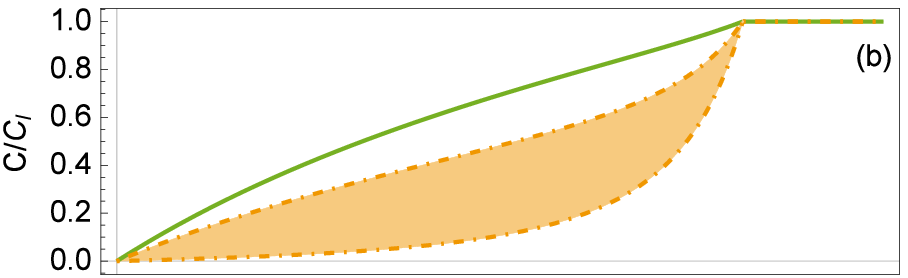}\
\includegraphics[width=8.05cm]{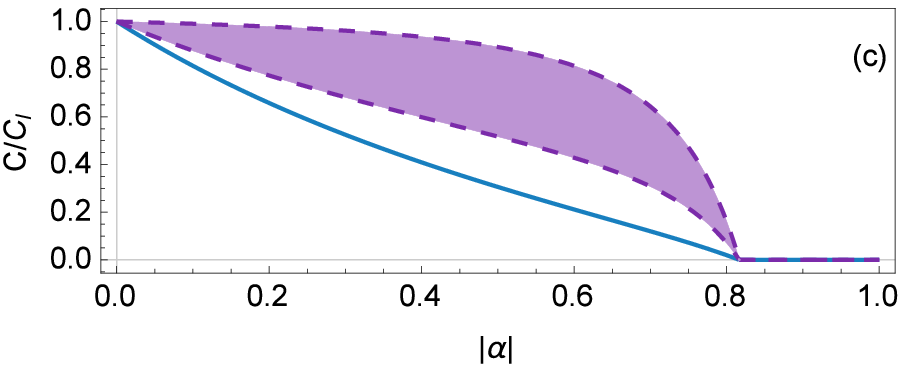} \\
\caption{(Color online) Various intrinsic coherence contribution in (a) and their proportions in the total amount in (b) and (c) as functions of the overlap $|\alpha|$ when $|\alpha_c|=0.8$ and $p_+=0.4$. In (a), the solid line shows the total intrinsic coherence, the dashed lines are for $\mathcal{C}_{S:CA}$ and $\mathcal{C}_{S:C}$ with $\mathcal{C}_{S:CA} \geq \mathcal{C}_{S:C}$, and the dot-dashed lines are for $\mathcal{C}_{A:SC}$ and $\mathcal{C}_{C:A}$ with $\mathcal{C}_{A:SC} \geq \mathcal{C}_{C:A}$, for $\gamma=\pi/2$. In (b) [(c)], the solid line shows ${\mathcal{C}_{A:SC}}/{\mathcal{C}_{I}}$ (${\mathcal{C}_{S:C}}/{\mathcal{C}_{I}}$), and the colored areas shows the region of ${\mathcal{C}_{C:A}}/{\mathcal{C}_{I}}$ (${\mathcal{C}_{S:CA}}/{\mathcal{C}_{I}}$).
} \label{fig3}
\end{figure}

From Fig. \ref{fig3}(a) one can notice the trade-off relation between $\mathcal{C}_{S:CA}$ and $\mathcal{C}_{C:A}$, as well as the one between $\mathcal{C}_{A:SC}$ and $\mathcal{C}_{S:C}$, with their sum being $\mathcal{C}_{I}=\mathcal{C}_{C:AS}$.
We also plot \rvs{their} proportions in the total intrinsic coherence in Fig. \ref{fig3}(b) and (c).
The region of $|\alpha|$ is divided into two parts according to the optimal conditions of $P_{\rm suc,max}$.
When $|\alpha| < \tilde{\alpha}$, ${\mathcal{C}_{A:SC}}/{\mathcal{C}_{I}}$ is a monotonous increasing function of $|\alpha|$ and ${\mathcal{C}_{S:C}}/{\mathcal{C}_{I}}$ is a decreasing one, both of which are independent of the Berry phase $\gamma$.
However, the other pair, $\mathcal{C}_{S:CA}/{\mathcal{C}_{I}}$ and $\mathcal{C}_{C:A}/{\mathcal{C}_{I}}$, depends on the value of $\gamma$.
Considering the relations $\mathcal{C}_{S:CA} = \mathcal{C}_{S:C} + \mathcal{C}_{g}$ and $\mathcal{C}_{C:A}=\mathcal{C}_{A:SC}-  \mathcal{C}_{g}$, one can determine the shaded areas in Fig. \ref{fig3}(b) and (c) by searching for the extreme points of $\mathcal{C}_{g}/\mathcal{C}_{I}$.
For fixed amounts of $p_+$, $|\alpha|$ and $|\alpha_c|$, the minimum of $\mathcal{C}_{S:CA}/{\mathcal{C}_{I}}$ occurs at $\gamma=0$ and $\pi$, while the maximum at $\cos\gamma=  -  |\alpha_c|$.
When $|\alpha| \geq \tilde{\alpha}$,  the separability condition of $\rho_{SA}$ is $\beta=\pi/2$. And the three-qubit state is a direct-product state as
\begin{equation}
|\Gamma\rangle=|1\rangle_s \otimes (u_{a} \otimes \openone_c) |\chi\rangle_{ac}
\end{equation}
where $|\chi\rangle_{ac}$ is the initial state $|\chi\rangle$ with the qubit $S$ replaced by the auxiliary system $A$, and $u_{a} $ is the local unitary operator for the ancilla satisfying $u_{a}  |\xi\rangle_a = \alpha_+ |1\rangle_a$ and $u_{a}  |\bar{\xi}\rangle_a = \bar{\alpha}_- |0\rangle_a + \alpha_- |1\rangle_a$.
Hence, the initial intrinsic coherence is completely transformed to the one between $S$ and $C$.
That is, $\mathcal{C}_{C:A}=\mathcal{C}_{A:SC}=\mathcal{C}_{C:AS}=\mathcal{C}_{I}$ and $\mathcal{C}_{S:CA}=\mathcal{C}_{S:C}=\mathcal{C}_{g}=0$.
These results indicate that the procedure of optimal USSD requires the transformations of intrinsic coherence from $\mathcal{C}_{S:C}$ to $\mathcal{C}_{A:SC}$, and from $\mathcal{C}_{S:CA}$ to $\mathcal{C}_{C:A}$.
The transformation rates increase with the inner product $|\alpha|$, and reach $1$ when $|\alpha| \geq \tilde{\alpha}$.
The transformation from $\mathcal{C}_{S:CA}$ to $\mathcal{C}_{C:A}$ depends on the Berry phase, as the genuine tripartite coherence is generated by the joint unitary transformation $U_{SA}$.

\section{Probabilistic teleportation}\label{Telepor}

USSD is an important step in the protocol of probabilistic teleportation \cite{ConcluTeleport,PRA2015RoaTele}.
To illustrate the scheme, we denote the Pauli operators as $\sigma_x^{x,y,z}$ and the Bell states as $|\psi_{\pm}\rangle_{xy}=(|0\rangle_x |0\rangle_y \pm |1\rangle_x |1\rangle_y  )/\sqrt{2}$ and  $|\phi_{\pm}\rangle_{xy}=(|0\rangle_x |1\rangle_y \pm |1\rangle_x |0\rangle_y  )/\sqrt{2}$, with the subscripts indicating different qubits.
Let us consider the sender Alice and the receiver Bob shearing a partially entangled pure state of two qubits $B$  and $C$ as
\begin{equation}\label{quantumchannel}
|\psi\rangle_{bc}=(M_b\otimes\openone_c)|\psi_+\rangle_{bc},
\end{equation}
where $M_b= \cos \rho \openone_b + \sin \rho \sigma_b^z$ and $\rho \in [0,\pi/4]$.
Its tangle is ${\rm Con}^2(\psi)= \cos^2 2\rho$.
We remark that, the results in this part can be easily generalized to the case with an arbitrary two-qubit pure state, by acting on the state  (\ref{quantumchannel}) and the following formulas with two local unitary operators of $B$ and $C$.
That is, \rvs{the following results about} success probability and intrinsic coherence are universal.

The aim of teleportation is to send an unknown state $|\varphi\rangle_s=\cos \frac{\mu}{2}|0\rangle_s+\sin \frac{\mu}{2}e^{i\nu}|1\rangle_s$ of qubit $S$ in Alice hand to the Bob's qubit $C$.
One can find that the three-qubit state $|\varphi\rangle_s \otimes|\psi\rangle_{bc}$ satisfies the following identity,
\begin{eqnarray}
|\varphi\rangle_s\otimes|\psi\rangle_{bc}  \!\!\!   &=& \!\!\!   \frac{1}{2}(|\psi'_+\rangle_{sb}|\varphi\rangle_c+ |\psi'_-\rangle_{sb} \sigma_c^z|\varphi\rangle_c \ \ \ \  \ \ \ \ \ \ \ \ \ \ \ \ \  \nonumber \\
 &\ &\    +|\phi'_+\rangle_{sb} \sigma_c^x|\varphi\rangle_c+|\phi'_-\rangle_{sb} \sigma_c^x\sigma_c^z|\varphi\rangle_c),
\end{eqnarray}
where $|\psi'_{\pm}\rangle_{sb}=(\openone_s \otimes M_b)|\psi_{\pm}\rangle_{sb}$, $|\phi'_{\pm}\rangle_{sb}=(\openone_s \otimes M_b)|\phi_{\pm}\rangle_{sb}$, and $|\varphi\rangle_c=\cos \frac{\mu}{2}|0\rangle_c+\sin \frac{\mu}{2}e^{i\nu}|1\rangle_c$.
To transform the states $|\psi'_{\pm}\rangle_{sb}$ and $|\phi'_{\pm}\rangle_{sb}$ into a separable form, Alice applies a controlled-NOT gate on her two qubits, in which $S$ controls $B$, followed by a Hadamard gate on qubit $S$.
Set $|\xi_\pm\rangle_s = \cos \rho |0\rangle_s \pm \sin\rho |1\rangle_s$, $|\bar{\xi}_\pm\rangle_s=\sigma_s^x|\xi_\pm\rangle_s$, and  $|\chi_\pm \rangle_{sc}$ to be two normalized states of $SC$ as
\begin{subequations}\label{ChiPM}
\begin{align}
|\chi_+ \rangle_{sc} \propto |\xi_+\rangle_s \otimes |\varphi\rangle_c  +|\bar{\xi}_+\rangle_{s}\otimes \sigma_c^z|\varphi\rangle_c,\ \ \ \ \ \ \  \\
|\chi_- \rangle_{sc} \propto |\xi_-\rangle_s \otimes \sigma_c^x |\varphi\rangle_c  +|\bar{\xi}_-\rangle_{s} \otimes \sigma_c^x \sigma_c^z|\varphi\rangle_c.
\end{align}
\end{subequations}
The state of the whole three-qubit system can be written as
\begin{equation}
|\Omega\rangle=\sqrt{P_+} |0\rangle_b \otimes |\chi_+ \rangle_{sc}+\sqrt{P_-}|1\rangle_b \otimes |\chi_-\rangle_{sc},
\end{equation}
where $P_\pm= {(1   \pm \sin 2\rho\cos\mu)}/{2} $.
Then, Alice performs a von Neumann measurement on qubit $B$, and collapses $SC$ onto one of the two possible states $|\chi_\pm \rangle_{sc}$ with the probabilities $P_\pm$.
Depending on the outcome, Alice performs the optimal USSD on qubit $S$ entangled with $C$.
If discrimination succeeds, she sends the results of qubits $B$ and $S$ to Bob over a classical communication channel.
Basing on these results, Bob applies one of four unitary operations $\{\openone_c, \sigma_c^x, i\sigma_c^y, \sigma_c^z\}$ on qubit $C$ to transform the state of qubit $C$ into  $|\varphi\rangle_c$, completing the teleportation.
Otherwise, the teleportation fails when Alice fails in the USSD.

For entangled state $|\chi_\pm \rangle_{sc}$, one can obtain the parameters in \rvs{USSD} as $p_+=1/2$, $\alpha=\pm \sin 2\rho$, $\alpha_c=\cos \mu$, and consequently the Berry phase $\gamma=0$ or $\pi$.
The total success probability of the teleportation can be calculated to be
\begin{eqnarray}
P_{\rm suc,max}^{\rm Tel} &=& P_+ P_{\rm suc,max}( \chi_+  ) + P_- P_{\rm suc,max}( \chi_-  ), \nonumber\\
&=& 1- \sin 2\rho,
\end{eqnarray}
where $P_{\rm suc,max}( \chi_\pm)$ stands for the optimal success probability of the USSD for entangled state $|\chi_\pm \rangle_{sc}$.

\begin{figure}
\centering
\includegraphics[width=8cm]{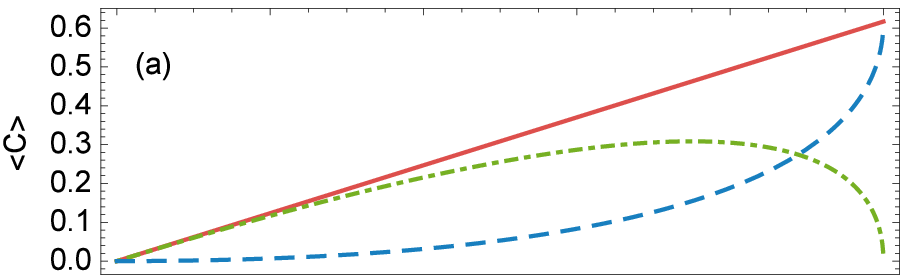} \
\includegraphics[width=8.02cm]{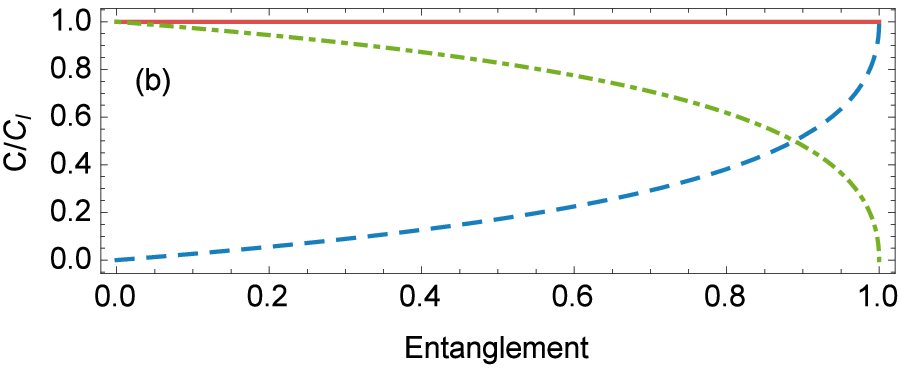} \\
\caption{(Color online) Square mean root of various intrinsic coherence contributions in (a) and their proportions in the total amount in (b) as functions of the entanglement (tangle) of the quantum channel in probabilistic teleportation. The solid lines show the total intrinsic coherence, the dashed lines are for $\mathcal{C}_{S:C}$, and the dot-dashed lines are for $\mathcal{C}_{A:SC}$.
} \label{fig4}
\end{figure}

In both cases, the condition for the absence of $\mathcal{C}_{A:S}$ is $|\eta\rangle_s=(|0\rangle_s+|1\rangle_s)/\sqrt{2}$, which is independent of  the state to send.
In other words, Alice can always perform the optimal USSD without the knowledge of the state $|\varphi\rangle_c$, with the aid of the total intrinsic coherence provided by the state $|\chi_\pm \rangle_{sc}$.
The various contributions of intrinsic coherence in (\ref{CIvalue}) become
\begin{eqnarray}\label{CIvaluePT}
&&\mathcal{C}_I^\pm=\frac{\sin^2\mu}{4 P_{\pm}^2}  \cos^2 2\rho, \nonumber  \\
&&\mathcal{C}_{A:SC}^\pm=\mathcal{C}_{g}^\pm=\frac{\sin^2\mu}{ 4 P_{\pm}^2}    2(\sin2\rho-\sin^2{2\rho}),
\end{eqnarray}
where the signs $\pm$ in superscripts and subscripts correspond to the one in $|\chi_\pm \rangle_{sc}$.
The transformation from $\mathcal{C}_{S:CA}$ to $\mathcal{C}_{C:A}$ does not take place, as the only coherence between $A$ and $SC$ originates from the genuine tripartite coherence.
To analyze the roles of these various coherence contributions in the protocol of probabilistic teleportation, we define their \emph{square mean root}
\begin{eqnarray}
<\!\! \mathcal{C}\!\! > = \biggr[ \int^{2\pi}_0 \!\!\!\!\!\int^\pi_0 \!\!\! \frac{1}{4\pi} \bigr(P_+ \sqrt{\mathcal{C}^+} + P_- \sqrt{\mathcal{C}^-} \bigr)  \sin\mu  d\mu d\nu \biggr]^2,
\end{eqnarray}
where the double integral denotes an average over the states to be sent uniformly distributed on the Bloch sphere.
In addition, the proportions of intrinsic coherence in (\ref{CIvaluePT}) in the total amount do not depend on the outcome of the measurement on $B$ or the state to be sent.
In Fig. \ref{fig4}, one can notice that, the square mean root of total intrinsic coherence increases with the entanglement of the quantum channel for teleportation.
As the state $|\psi\rangle_{bc}$ turns from the maximal entangled state into a separable one, the transformation rate from $\mathcal{C}_{S:C}$ to $\mathcal{C}_{A:SC}$ increases from $0$ to $1$.
Hence, the total intrinsic coherence is enhanced by the quantum feature of the entangled channel.
Moreover, the proportion $\mathcal{C}_{A:SC}/\mathcal{C}_{I}$ is related to the importance of the auxiliary qubit in this protocol, which decrease with the entanglement of the channel.

\section{\rvs{Summary}}\label{Summ}

We study the procedure of USSD, in which a qubit $S$ is entangled with another one $C$, assisted by an auxiliary system $A$.
\rvs{
The initial-state preparation of $S$ is \emph{self-contained}, as the randomness of its two states arises from the entanglement with its environment $C$.
In other words, we avoid the classical probabilities and take all degrees of freedom influencing the quantumness of this process into account.
Further, unambiguous state discrimination with \emph{a priori} classical probabilities \cite{roa2011dissonance,PhysRevA.85.022328,zhang2012assisted}, can be considered as a special case our procedure.
}

The \rvs{ intrinsic coherence and its distribution } as quantum features in such tripartite process are the focus of this article.
\rvs{
The optimal success probability to recognize the two states of qubit $S$ is enhanced by its initial intrinsic coherence with $C$.
However, no extra intrinsic coherence beyond the initial one is required in the optimal process,
as one can keep $\rho_{SA}$ as a separable state by adjusting the joint unitary $U_{SA}$ without affecting the success probability.
Furthermore, the transformations of intrinsic coherence, from $\mathcal{C}_{S:C}$ to $\mathcal{C}_{A:SC}$ and from $\mathcal{C}_{S:CA}$ to $\mathcal{C}_{C:A}$,
always occur and increase with the overlap between the two states to discriminate.
They become complete simultaneously in the case in which $\rho_{SA}$ is a classical state.
The genuine tripartite coherence generated in this \rvs{process} is closely related to the Berry phase $\gamma$, and influences the transformation from $\mathcal{C}_{S:CA}$ to $\mathcal{C}_{C:A}$.
 }

We also investigate the protocol of probabilistic teleportation as an application of the procedure of USSD.
The entanglement of the channel increases the total intrinsic coherence in optimal USSD, but decreases the proportion $\mathcal{C}_{A:SC}/\mathcal{C}_{I}$ corresponding to the importance of the ancilla in the procedure.


\acknowledgments
This work is supported by the NSF of China (Grant No. 11675119, No. 11575125 and No. 11105097).

\bibliography{USSD_EPL}

\end{document}